\documentclass[aps,prb,twocolumn,superscriptaddress,longbibliography]{revtex4-1}
\usepackage{graphicx} 
\usepackage{graphicx}
\usepackage{graphics}
\usepackage{amsmath}
\usepackage{amsthm} %
\usepackage{amssymb}%
\usepackage{amsfonts}
\usepackage{amssymb}
\usepackage{epstopdf}
\usepackage{makeidx}
\usepackage{epsfig}
\usepackage{amsfonts}
\usepackage{bm}
\usepackage{color}
\usepackage{xcolor}
\usepackage{bbold}
\usepackage{afterpage}
\usepackage{empheq}
\usepackage{hyperref}
\usepackage{booktabs}
\usepackage{subfigure}
\usepackage{makecell}
\usepackage{mathtools}
\usepackage{pgfplots}
\usepackage{bm}
\pgfplotsset{compat=1.18}
\usepackage{tikz}
\usetikzlibrary{arrows}
\newcommand{\be}{\begin{equation}}
\newcommand{\ee}{\end{equation}}
\newcommand{\ba}{\begin{align}}
\newcommand{\ea}{\end{align}}

\begin{document}
\title{Intrinsic Orbital Hall Effect in Degenerate Spin-3/2 Systems driven by the quantum metric}

\author{Rhonald Burgos Atencia}

\affiliation{Dipartimento di Fisica "E. R. Caianiello", Universit\`a di Salerno, IT-84084 Fisciano (SA), Italy}

\begin{abstract}
We show that in rotationally invariant spin-$3/2$ systems the orbital Hall effect originates from interband matrix elements of the orbital magnetic moment. The resulting Hall response is governed by the quantum metric, rather than by the Berry curvature, revealing a purely geometric transport mechanism. Conventional intraband contributions associated with the orbital magnetic moment and Berry curvature are shown to vanish identically in degenerate systems. These results identify the quantum metric as the key geometric quantity controlling orbital Hall transport in degenerate multiband systems.
\end{abstract}

\maketitle

\textit{Introduction.–}
The intrinsic orbital magnetic moment (OMM) of Bloch electrons is a central quantity governing orbital-related nonequilibrium phenomena.
Within a semiclassical picture, it is interpreted as the self-rotation of an electronic wave packet about its center of mass \cite{Chang2008,XiaoDi2010,BurgosAtencia2024}.
Of particular importance is the flow of this orbital moment in the presence of external driving fields, which gives rise to orbitally polarized currents \cite{Bernevig2005,Canonico2020,Go2021,CysneTarik2022,PezoArmando2022,PezoArmando2023II,CostaMarcio2023,Barbosa2024,CysneTarik2024,Cullen2025,HongLiu2025,Ando2025,WangPing2025}.

Within semiclassical theory, the OMM is inherently a band-diagonal quantity, i.e., it possesses only intraband matrix elements.
Such intraband contributions have been shown to generate an orbital Hall response in a variety of systems \cite{Bhowal2021,LiuHong2024,Tarik2024,Tang2024}.
In two-band models, the OMM can be expressed directly in terms of the Berry curvature; however, no such direct relation exists in multiband systems.
As a consequence, the OMM may remain finite even in situations where the Berry curvature vanishes, reflecting the fact that it generally involves interband contributions beyond those captured by the Berry curvature alone.
Moreover, in multiband systems the Berry curvature may change sign between states of opposite helicity while the OMM retains the same sign, indicating that the two quantities encode distinct aspects of band geometry.
Such behavior has been reported, for example, in systems hosting multifold fermions \cite{Flicker2018}, where the Berry curvature of the 
$m=0$ band of an SU(3) state vanishes but the OMM is still finite.

This picture applies to nondegenerate bands or to systems with isolated degeneracy points in the Brillouin zone.
When bands are degenerate at every momentum, however, both the geometric structure and the transport theory change qualitatively and quantitatively.
In this case, the intrinsic geometric properties of the band structure acquire a fundamentally matrix character that can be identified with a 
non-Abelian structure \cite{Wilczek1984} and which profoundly modifies the structure of transport theory, particularly the role played by the semiclassical intraband OMM.
Such a situation generically arises in three-dimensional systems with time-reversal symmetry $\Theta$ and inversion symmetry $\mathcal{I}$ \cite{Haldane2004}.
A paradigmatic realization is provided by rotationally invariant spin-$3/2$ systems described by the Luttinger Hamiltonian \cite{Luttinger1955,Luttinger1956,Murakami2004}, which constitute the focus of the present work.

A direct consequence of the combined presence of time-reversal $\Theta$ and inversion $\mathcal{I}$ symmetries is the complete suppression of the Berry curvature at each $\bm{k}$-point in the Brillouin zone \cite{XiaoDi2010}.
This occurs because, within a degenerate band, physical eigenstates are arbitrary superpositions of orthogonal basis states spanning the degenerate manifold.
As a result, a unique physical state cannot be specified by a single eigenvector and must instead be described by a density matrix, reflecting the freedom to form coherent superpositions within the degenerate subspace.
Although a matrix-valued Berry curvature can be formally defined within the degenerate manifold, the physically observable Berry curvature relevant for transport is obtained as the trace over the degenerate subspace and, owing to the presence of states with opposite helicity, vanishes identically.

A closely related structure emerges for the OMM in degenerate bands.
Although a matrix-valued OMM is well defined, its physically relevant intraband component is given by the trace over the degenerate subspace and is 
therefore directly tied to the Berry curvature in the present two-band Luttinger model.
with two pairs of degenerate bands due to 
combined inversion and time-reversal symmetry, this trace vanishes identically, implying that the semiclassical OMM contribution to the orbital dynamics is strictly zero.
Remarkably, however, the interband components of the OMM survive and generate an orbital Hall response whose geometric structure has so far remained unexplored and which constitutes the only intrinsic contribution that persists in fully degenerate bands.
Importantly, this response is not controlled by the Berry curvature—which vanishes—but by the quantum metric, namely the symmetric part of the quantum geometric tensor \cite{Provost1980,Jiang2025,YuJiabin2025}.
This contribution has no semiclassical counterpart and is intrinsically quantum mechanical.

In this work, we investigate the orbital magnetic moment generated in the Luttinger Hamiltonian and elucidate its structure, microscopic origin, and consequences for orbital transport.
Our strategy is to exploit the internal rotational symmetry of an underlying nondegenerate, time-reversal–symmetric manifold as an auxiliary tool to access the more intricate degenerate spectrum protected by both time-reversal and inversion symmetries.
As we explicitly show, the geometric structures of the two descriptions are closely connected, which allows for a fully analytical treatment of the orbital response in rotationally invariant spin-$3/2$ systems.

\textit{Theoretical framework.–} 
We consider the spherical Luttinger model described by the Hamiltonian \cite{Murakami2004}
\begin{equation}
\label{Eq:spericalham}
H=
\frac{\hbar^2}{2m_0}\left[\left(\gamma_1+\frac{5}{2}\gamma_2 \right)k^2
-2\gamma_2(\bm k\cdot \bm S)^2\right],
\end{equation}
where $m_0$ is the electron mass, $\gamma_{1,2}$ are material-dependent parameters, and $\bm S$ denotes
the four-dimensional representation of the $SU(2)$ algebra, satisfying the commutation relations
$[S_i,S_j]=i\epsilon_{ijl}S_l$.

To elucidate the band structure of Eq.~\eqref{Eq:spericalham} and its underlying intrinsic geometry, it is useful
to first consider the linear operator $H_L=\bm k\cdot \bm S$.
Its eigenstates are simultaneously eigenstates of the quadratic operator
$H_Q=(\bm k\cdot \bm S)^2$, with eigenvalues
$\varepsilon^L_{m,\bm k}=m|\bm k|$ and $\varepsilon^{Q}_{m,\bm k}=m^2|\bm k|^2$,
respectively, where $m=\pm 3/2,\pm 1/2$.
As a consequence, the quadratic Hamiltonian depends only on $m^2$ and
does not distinguish the helicity
$\hat{h}_0=\hat{\bm k}\cdot \bm S$.
This leads to a twofold degeneracy at each $\bm k$-point of the quadratic spectrum,
reflecting the underlying helicity symmetry of the Luttinger Hamiltonian.
This twofold degeneracy has direct and nontrivial consequences for orbital-moment-related responses.
Because the quadratic Hamiltonian describes a degenerate band manifold at each $\bm k$-point, the OMM 
must be treated as a matrix-valued operator acting within the degenerate subspace.
As a result, the conventional semiclassical description based solely on band-diagonal (intraband) contributions breaks down, and interband matrix elements become essential for capturing the intrinsic orbital response.

To proceed with the calculation of the OMM, it is convenient to first analyze the linear Hamiltonian $H_L=\bm k\cdot\bm S$.
Although the physical spectrum of the Luttinger model is governed by the quadratic operator $H_Q=(\bm k\cdot\bm S)^2$, the two Hamiltonians share closely related quantum-geometric structures.
The linear Hamiltonian $H_L$ preserves time-reversal symmetry but breaks inversion symmetry, lifting the helicity degeneracy and allowing the OMM to be computed within a nondegenerate framework.
Our strategy is to exploit this simplification: we first solve the problem associated with $H_L$, and subsequently reconstruct from it the full band geometry of the degenerate operator $H_Q$.
This approach provides a technically simpler and physically transparent route to the evaluation of the OMM, while retaining all the information required to describe the degenerate problem.

The eigenstates of the linear Hamiltonian $H_L=\bm k\cdot\bm S$ are labeled by the band index $m$ and momentum $\bm k$, and satisfy
$H_L |u_{m,\bm k}\rangle = \varepsilon^{L}_{m,\bm k} |u_{m,\bm k}\rangle$.
For later convenience, we distinguish states with positive and negative band index by introducing the notation $|u^{\uparrow}_{m,\bm k}\rangle$ and
$|u^{\downarrow}_{m,\bm k}\rangle$, corresponding to $+m$ and $-m$, respectively.
As will be shown below, these states naturally combine into degenerate doublets when the quadratic Hamiltonian is considered.
The OMM of the degenerate bands can then be expressed entirely in terms of matrix elements evaluated within the linear problem.

\begin{figure*}[t]
\centering
\includegraphics[width=\columnwidth]{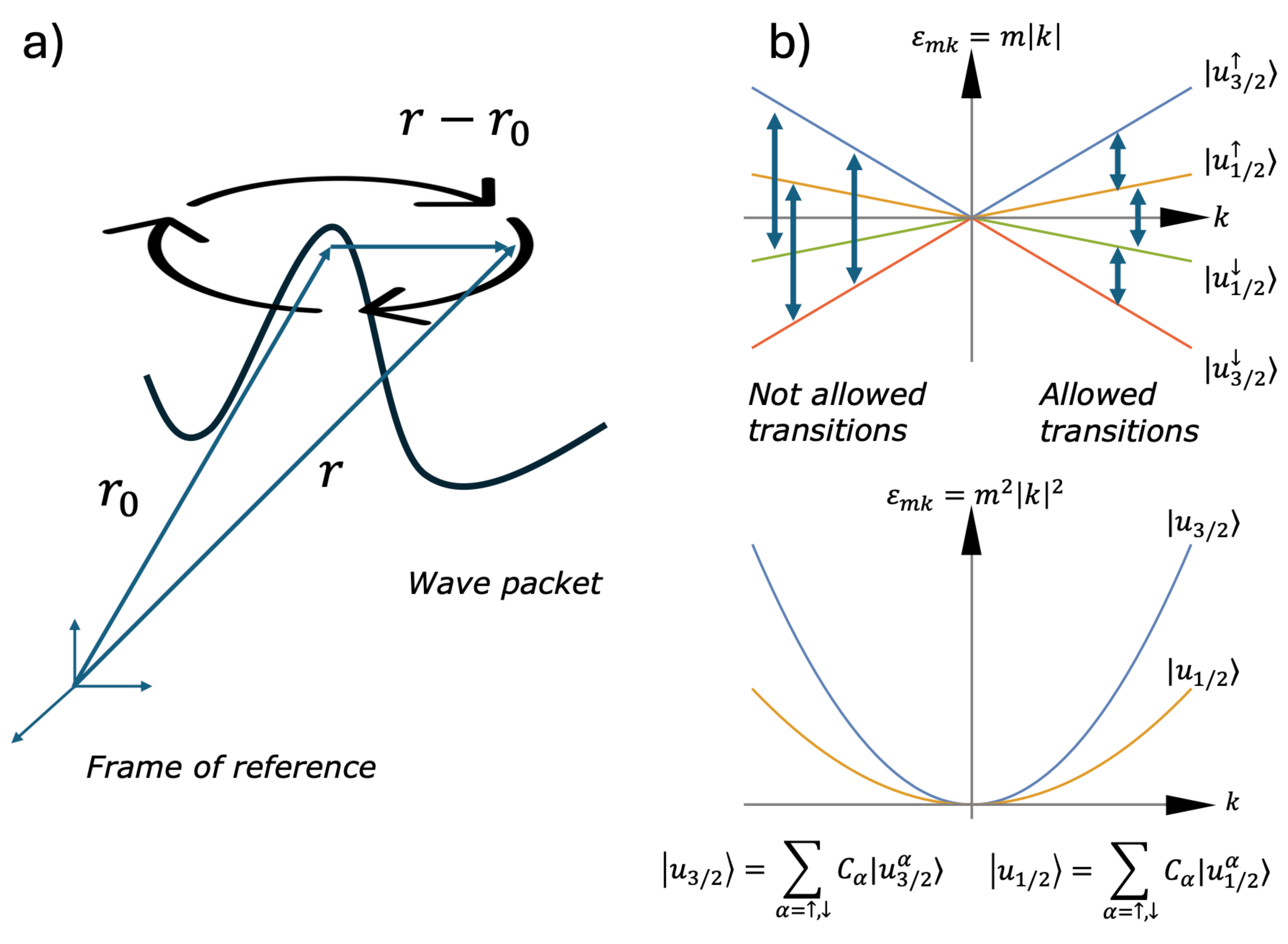} 
\includegraphics[width=\columnwidth]{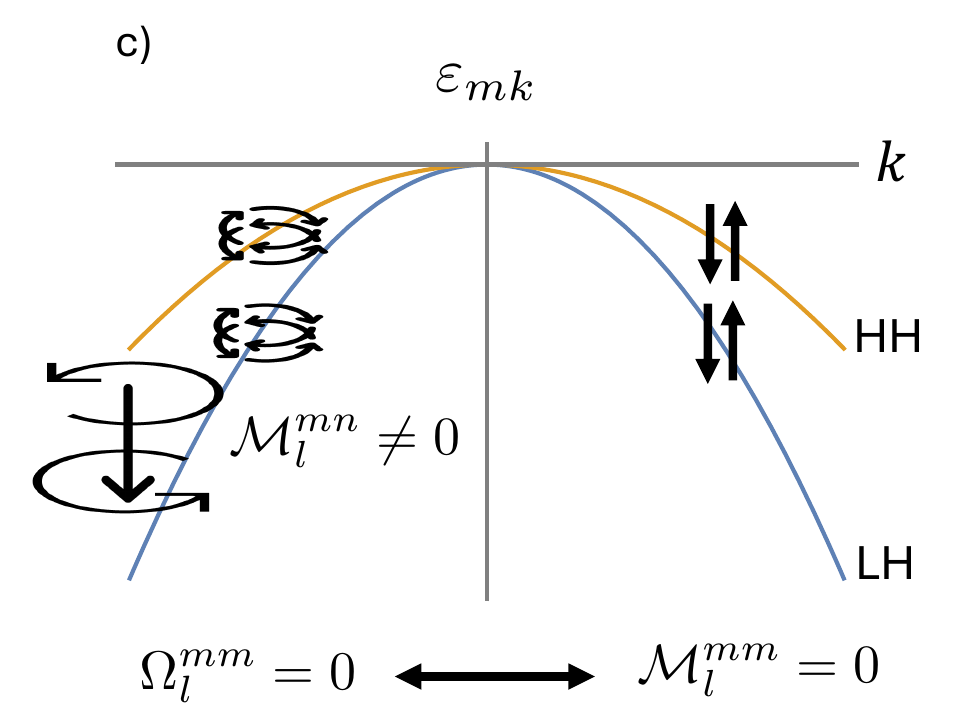}
\caption{
(a) The classical orbital angular momentum depends on the choice of reference frame.
While $\bm L=\bm r \times \bm v$ is origin dependent, the combination
$\bm L=(\bm r-\bm r_0)\times \bm v$ is defined with respect to the center of the wave packet,
as employed in the semiclassical description of the orbital magnetic moment.
(b) Schematic dispersion relations of the four-dimensional $SU(2)$ model for the
linear and quadratic Hamiltonians.
The upper panel illustrates the allowed and forbidden interband transitions
associated with the Berry connection, highlighting the nonvanishing matrix elements of the vector Berry connection
$\bm{\mathcal A}^{L}= i R^{\dagger}(\theta,\phi)\nabla_{\bm k} R(\theta,\phi)$.
The lower panel shows the quadratic spectrum, whose eigenstates are linear combinations
of the underlying linear-$SU(2)$ states with opposite helicity. 
c) The net intraband orbital magnetization vanishes in each band as a direct consequence of the geometry associated with the twofold-degenerate spectrum. 
The heavy-hole band corresponds to $|m|=3/2$ and is formed by the degenerate manifold $\{ |u^{\uparrow}_{3/2} \rangle, |u^{\downarrow}_{3/2} \rangle \}$, 
whereas the light-hole band corresponds to $|m|=1/2$ and is formed by the manifold $\{ |u^{\uparrow}_{1/2} \rangle, |u^{\downarrow}_{1/2} \rangle \}$. 
The cancellation follows from the exact compensation between opposite-helicity contributions within each degenerate subspace. 
This illustrates why no intraband orbital Hall response can arise in the degenerate quadratic spectrum. }
\label{fig:LM}
\end{figure*}

\textit{Intraband OMM of the linear Hamiltonian.–}
The orbital angular momentum is classically defined as
$\bm L = \bm r \times \bm v$.
Already at the classical level, this definition involves a subtlety:
the orbital angular momentum depends on the choice of reference frame,
since the position vector $\bm r$ is defined relative to an arbitrary origin
(see Fig.~\eqref{fig:LM}).
Within the semiclassical theory of transport, a frame-invariant physical
definition is obtained by choosing the center of mass of the wave packet
as the reference point, leading to
$\bm L = (\bm r - \bm r_0) \times \bm v$,
where $\bm r_0$ fixes the origin of the coordinate system \cite{Sundaram1999}.
This classical ambiguity foreshadows the geometric nature of orbital
magnetism in quantum systems.

In quantum mechanics, the OMM operator is defined as \cite{SongJustin2019}
\begin{equation}
\label{Eq:OMMeq}
\hat{\mathcal{M}}=-\frac{e}{4}(\hat{\bm r} \times \hat{\bm v} - \hat{\bm v} \times \hat{\bm r}).
\end{equation}

Its matrix elements in the Bloch eigenstate basis
coincide with the semiclassical expression \cite{Sundaram1999} provided the intraband Berry connection
is set to describe the center of the wave-packet. This choice corresponds to the parallel-transport gauge \cite{Pozo2023}, which we adopt throughout. 
Within this framework, the intrinsic OMM associated with a non-degenerate Bloch band $m$
takes the well-known form \cite{SongJustin2019}
\begin{align}
\label{Eq:OMMSM}
\mathcal{M}^{(L)mm}_{l}
&= -\frac{e}{2\hbar}\epsilon_{lij} {\rm Im} \sum_{n\neq m}
(\varepsilon_{m,\bm k} - \varepsilon_{n,\bm k})
\mathcal{A}^{(L)mn}_{i}\mathcal{A}^{(L)nm}_{j},
\end{align}
where $\mathcal{A}^{(L)mn}_i = i\langle u_{m,\bm k}|\partial_{k_i}|u_{n,\bm k}\rangle$ denotes the Berry connection related to the linear Hamiltonian $H_L$. Exploiting the rotational symmetry of the system, we choose the quantization axis along the helicity direction
$\hat{\bm h}_0=\bm k/|\bm k|$.
This is achieved by the rotation operator
$R(\theta,\phi)=e^{-i\phi S_z}e^{-i\theta S_y}$,
where $\phi=\tan^{-1}(k_y/k_x)$ and
$\theta=\cos^{-1}(k_z/|\bm k|)$.
Accordingly, the Bloch states can be written as
$|u_{m,\bm k}\rangle = R(\theta,\phi)\,|m\rangle$,
where $|m\rangle$ are eigenstates of $S_z$. Using this representation, the Berry connection acquires the compact form
$\mathcal{A}^{(L)}_i = \bm a_i \cdot \bm S$ with
$\bm a_i =
\bigl(
-\sin\theta\,\partial_{k_i}\phi,\;
\partial_{k_i}\theta,\;
\cos\theta\,\partial_{k_i}\phi
\bigr)$, which follows directly from $\bm{\mathcal A}^{(L)}= i R^{\dag}(\theta,\phi) \nabla_{\bm k} R(\theta,\phi)$.
Since the Berry connection is proportional to the spin operators, the structure of the orbital magnetic
moment is entirely dictated by the underlying SU(2) algebra.
As a result, only transitions between adjacent angular-momentum sectors, $\Delta m=\pm1$, contribute to the
sum over intermediate states in Eq.~\eqref{Eq:OMMSM}.
This selection rule follows from rotational symmetry and can be understood as a direct consequence of the
Wigner–Eckart theorem, since $\bm{\mathcal A}^{(L)}$ transforms as a rank-1 tensor \cite{BurgosPRB2026}.
In Fig.~\eqref{fig:LM} we illustrate the corresponding allowed transitions of the underlying linear spectrum.

Accordingly, the orbital magnetic moment can be decomposed as
$\mathcal{M}^{(L)mm}_{l} = \mathcal{M}^{(L)mm}_{l,+} + \mathcal{M}^{(L)mm}_{l,-}$, with
\begin{align}
\mathcal{M}^{(L)mm}_{l,\pm}
&= -\frac{e}{2\hbar}\epsilon_{lij}
(\varepsilon^{m}_{\bm k}-\varepsilon^{m\pm1}_{\bm k})
{\rm Im} \!\left(
\mathcal{A}^{(L)m,m\pm1}_{i}
\mathcal{A}^{(L)m\pm1,m}_{j}
\right).
\end{align}

To explicitly evaluate the matrix elements of the Berry connection, it is convenient to express the spin
operators in terms of ladder operators,
$S_{\pm}=S_x \pm i S_y$.
After straightforward algebra, we obtain \cite{BurgosPRB2026}
\begin{align}
\label{Eq:OMM}
\mathcal{M}^{(L)mm}_{l,\pm}
= - \frac{e}{8\hbar} \epsilon_{lij} \,
\hat{\bm h}\cdot ( \partial_i\hat{\bm h} \times \partial_j\hat{\bm h} ) \,
B^{\pm}(S,m),
\end{align}
with
\begin{align}
\label{Eq:factor}
B^{\pm}(S,m)= \mp (\varepsilon^{m}_{\bm k}-\varepsilon^{m\pm1}_{\bm k})
\,[S(S+1)-m(m\pm 1)],
\end{align}
where $S=3/2$ is the total angular momentum. Here $\mathcal{M}^{(L)mm}_{l}$ denotes the OMM associated with the linear Hamiltonian
$H_L$, which has a non-degenerate spectrum. For $S=1/2$, it is in agreement with Refs.~\cite{Yoda2018,ZhongShudan2016}.

\textit{Intraband OMM of the quadratic Hamiltonian.–}
Due to the twofold degeneracy of the quadratic Hamiltonian $H_Q$, its eigenstates
are not uniquely defined and can be expressed as arbitrary linear combinations
within the degenerate subspace (see Fig.~\eqref{fig:LM}).
A general eigenstate can therefore be written as
$|\phi_{m,\bm k}\rangle=\sum_{\alpha=\uparrow,\downarrow} C_\alpha\,|u^{\alpha}_{m,\bm k}\rangle$,
where $\{|u^{\uparrow}_{m,\bm k}\rangle,|u^{\downarrow}_{m,\bm k}\rangle\}$
form an orthonormal basis of the degenerate subspace and the coefficients satisfy normalization condition $|C_\uparrow|^2+|C_\downarrow|^2=1$. 
The notation $\uparrow, \downarrow$ is just to reflect the two dimensional structure of each degenerate manifold. 
Hence, it is equivalent to $\{ |u_{3/2} \rangle, |u_{-3/2} \rangle \}$ or $\{ |u_{1/2} \rangle, |u_{-1/2} \rangle \}$. For the quadratic 
Hamiltonian we reserve $m$ to be $m=3/2$ or $m=1/2$.

The Berry connection associated with the degenerate band is defined as
\begin{equation}
\mathcal{A}^{(Q)mn}_i
=
i\langle \phi_{m,\bm k}|\partial_{k_i}|\phi_{n,\bm k}\rangle,
\end{equation}
which can be written in compact form as
$\mathcal{A}^{(Q)mn}_i = \bm C_m^\dagger A^{mn}_i \bm C_n$,
where $\bm C_m=(C_\uparrow,C_\downarrow)^T$ and the Berry connection
matrix reads
\begin{equation}
\label{eq:matrixBC}
A^{mn}_{i}=
\begin{pmatrix}
i\langle u^{\uparrow}_{m,\bm k}|\partial_{k_i}|u^{\uparrow}_{n,\bm k}\rangle &
i\langle u^{\uparrow}_{m,\bm k}|\partial_{k_i}|u^{\downarrow}_{n,\bm k}\rangle \\
i\langle u^{\downarrow}_{m,\bm k}|\partial_{k_i}|u^{\uparrow}_{n,\bm k}\rangle &
i\langle u^{\downarrow}_{m,\bm k}|\partial_{k_i}|u^{\downarrow}_{n,\bm k}\rangle
\end{pmatrix}.
\end{equation}

Note that the scalar $\mathcal{A}^{(Q)mn}_i$ is defined with $|\phi_{n,\bm k}\rangle$ while the matrix-valued $A^{mn}_{i}$ is defined with 
the states in the helicity basis. The product of two Berry connections then reads
\begin{equation}
\mathcal A^{(Q)mn}_i \mathcal A^{(Q)nm}_j=\mathrm{tr} \left( \Gamma^m  A^{mn}_i \Gamma^n  A^{nm}_j\right),
\label{Eq:traceAA}
\end{equation}
where for each degenerate manifold we introduce the pure-state density matrix
$\Gamma^{s} \equiv \bm C_s \bm C^\dagger_s$, which encodes the internal polarization within the degenerate subspace. 
Any pure internal polarization within a degenerate manifold can be represented by
\begin{equation}
\Gamma^s_\zeta
=
\frac{1}{2}\bigl(I_s+\zeta\,\hat{\bm n}_s\cdot\bm\sigma\bigr),
\qquad
\zeta=\pm1,
\end{equation}
with $\hat{\bm n}$
a unit vector on the Bloch sphere. Here $\zeta=\pm1$ labels the two orthogonal internal polarizations of the degenerate manifold. 
Since the quadratic Hamiltonian possesses a twofold degeneracy, its eigenstates are not uniquely defined. 
In the absence of a symmetry-breaking perturbation, there is no preferred internal polarization within the degenerate subspace. In the present work, we represent each degenerate manifold by the basis-independent maximally mixed density matrix $\Gamma^s_\zeta=\frac{1}{2}I_s$, which provides a basis-independent description of the degenerate band.
This choice removes the arbitrariness associated with the internal gauge of the degenerate subspace while preserving all gauge-invariant observables.

The intraband OMM of the degenerate spectrum is defined using the scalar Berry connections of the quadratic Hamiltonian as
\begin{equation}
\label{Eq:OMMdegenerate}
\mathcal M^{(Q)mm}_l
=
-\frac{e}{2\hbar}\epsilon_{lij}\,
{\rm Im} \sum_{n\neq m}
\bigl(\varepsilon^{(Q)}_{m\bm k}-\varepsilon^{(Q)}_{n\bm k}\bigr)
\left(\mathcal A^{(Q)mn}_i\mathcal A^{(Q)nm}_j\right).
\end{equation}

We are now in a position to relate the band diagonal OMM of the quadratic Hamiltonian to that of the underlying linear Hamiltonian $H_L$, and to elucidate how the intrinsic 
symmetries of the system suppresses the intraband orbital response. To this end, we recall the definition of the Berry curvature,
$\Omega^{mm}_{ij}=-2 {\rm Im} \sum_{n\neq m}\mathcal{A}^{mn}_i \mathcal{A}^{nm}_j$,
which for the linear Hamiltonian takes the simple form
$\Omega^{mm}_{ij} = -m\, \hat{\bm h}\cdot ( \partial_i\hat{\bm h} \times \partial_j\hat{\bm h})$ \cite{BurgosPRB2026,WenKevin2025}. 
Since the eigenvalues of the quadratic Hamiltonian depend on $m^2$, its twofold-degenerate bands are formed by superpositions of states with opposite helicities.
Using the same definition of the Berry curvature but in terms of the scalar Berry connections of the quadratic Hamiltonian as
\begin{equation}
\label{Eq:BCdegenerate}
\Omega^{(Q)mm}_{ij}=-2{\rm Im} \sum_{n\neq m}\mathcal{A}^{(Q)mn}_i \mathcal{A}^{(Q)nm}_j,
\end{equation}
it follow that $\Omega^{(Q)mm}_{ij}=0$, namely, for the quadratic Luttinger Hamiltonian considered here, the Berry curvature constructed from the scalar Berry connections of the degenerate eigenstates vanishes identically. 
Mathematically, this result follows because the trace in Eq.~\eqref{Eq:traceAA} combines the contributions from the two opposite-helicity states forming each degenerate manifold. As shown previously for the linear Hamiltonian $H_L$ these two helicity sectors carry Berry curvatures of equal magnitude and opposite sign, which cancel exactly after taking the trace.

Returning to Eq.~\eqref{Eq:OMMdegenerate}, we note that, for the Luttinger Hamiltonian considered here, the intraband OMM has exactly the same mathematical structure as that of a non-degenerate two-band model. Once the index $m$ is identified with either the heavy hole (HH) or light hole (LH) degenerate manifold, there exists only a single interband transition, namely $\mathrm{HH}\leftrightarrow\mathrm{LH}$. Consequently, the summation over $n\neq m$ contains only one term, and Eq.~\eqref{Eq:OMMdegenerate} reduces directly to the definition of the Berry curvature Eq.~\eqref{Eq:BCdegenerate}.
Since $\Omega^{(Q)mm}_{ij}=0$, the intraband OMM also vanishes identically. We emphasize that this simplification is specific to the doubly degenerate two-band structure of the Luttinger Hamiltonian. In a multiband system containing more than two degenerate manifolds, the summation over $n\neq m$ involves additional interband transitions and the OMM can no longer be reduced solely to the Berry curvature. Therefore, the vanishing of the intraband OMM demonstrated here should not be regarded as a general consequence of $\mathcal{I}\Theta$-symmetry, but rather as a consequence of the particular band structure of the Luttinger model.

We conclude this subsection by emphasizing the relation between the linear and quadratic Hamiltonians and its implications for the geometric description. Since $H_Q = H_L^2$, both Hamiltonians share the same eigenstates, and the eigenbasis of $H_L$ provides a natural and convenient gauge choice within the degenerate subspace of $H_Q$. In the quadratic Hamiltonian, the energy depends on $m^2$, such that states with $m$ and $-m$ become degenerate. This allows one to reorganize the basis into two-dimensional subspaces, introducing an internal index $\uparrow,\downarrow$ to label the two orthogonal states within each degenerate manifold. 
We stress that this choice corresponds to a gauge fixing, and that the physical quantities considered in this work, being expressed in terms of traces or ensemble-averaged combinations, are invariant under local $U(2)$ rotations within the degenerate subspace.

\textit{Interband OMM and orbital Hall effect.–}
As established above, the conventional intraband (band-diagonal) contribution to the intrinsic OMM vanishes identically for the Luttinger Hamiltonian. This result has an important physical consequence. Since the orbital Hall response cannot originate from the diagonal orbital magnetic moment, any finite orbital transport must necessarily arise from the remaining matrix elements of the OMM operator. In fact, recent studies have emphasized that the complete orbital magnetic moment is fundamentally a matrix-valued operator and that its off-diagonal matrix elements play an essential role in describing nonequilibrium orbital transport phenomena such as the orbital Hall effect \cite{Cysne2026}. Motivated by this observation, we now investigate the interband OMM coupling the degenerate heavy-hole (HH) and light-hole (LH) manifolds. Unlike the intraband contribution, these interband matrix elements remain finite and constitute the physically relevant contribution responsible for the intrinsic orbital Hall response in the present model.

The corresponding matrix-valued interband contribution follows from Eq.~\eqref{Eq:OMMeq} and reads \cite{Cysne2026}
\begin{align}
\mathcal{M}^{mm'}_\gamma
=
-\frac{e}{4}\epsilon_{\gamma jl}
\big(
v^{mm}_{\bm k,l}
+
v^{m'm'}_{\bm k,l}
\big)
\,\mathcal{A}^{(Q)mm'}_{\bm k,j}.
\end{align}

\begin{figure}[t]
\centering
\includegraphics[width=0.48\columnwidth]{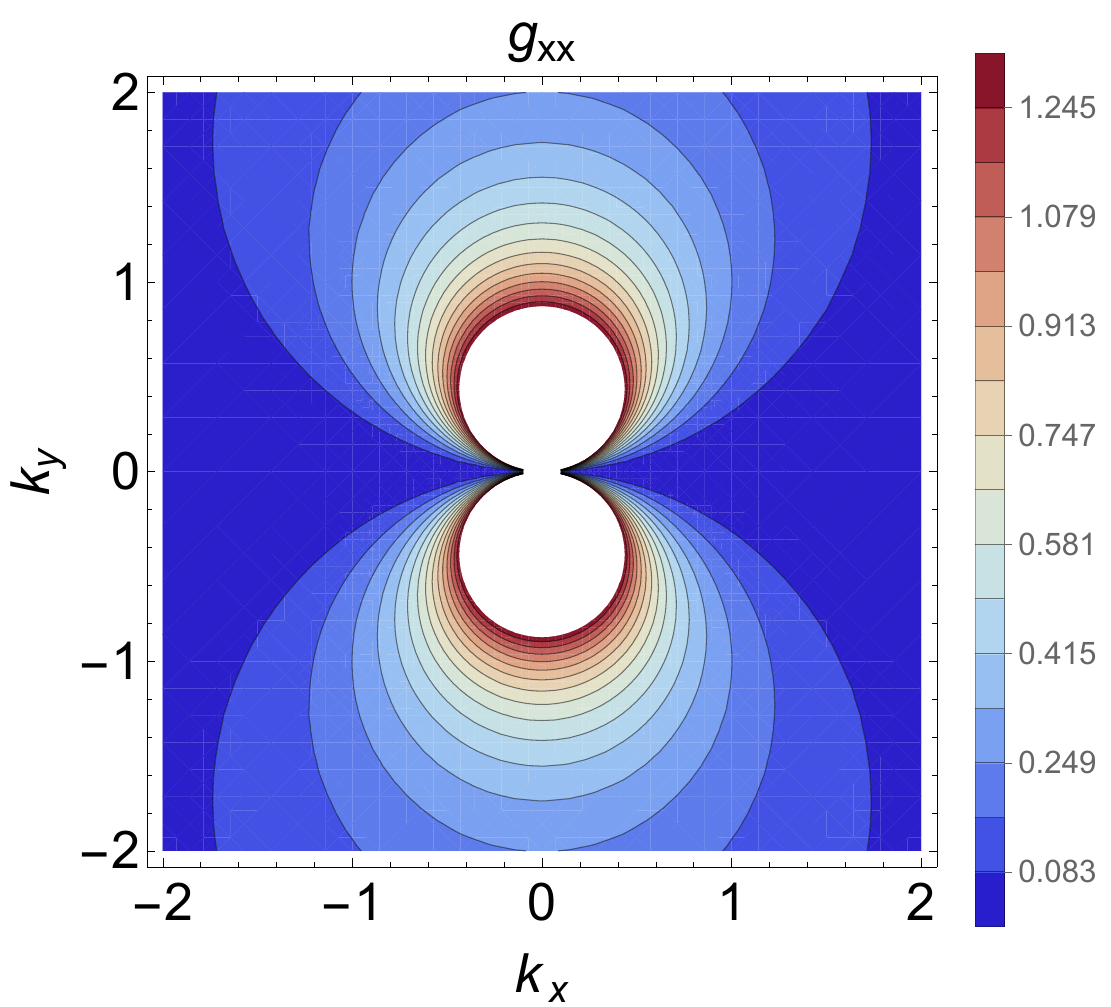}
\hfill
\includegraphics[width=0.48\columnwidth]{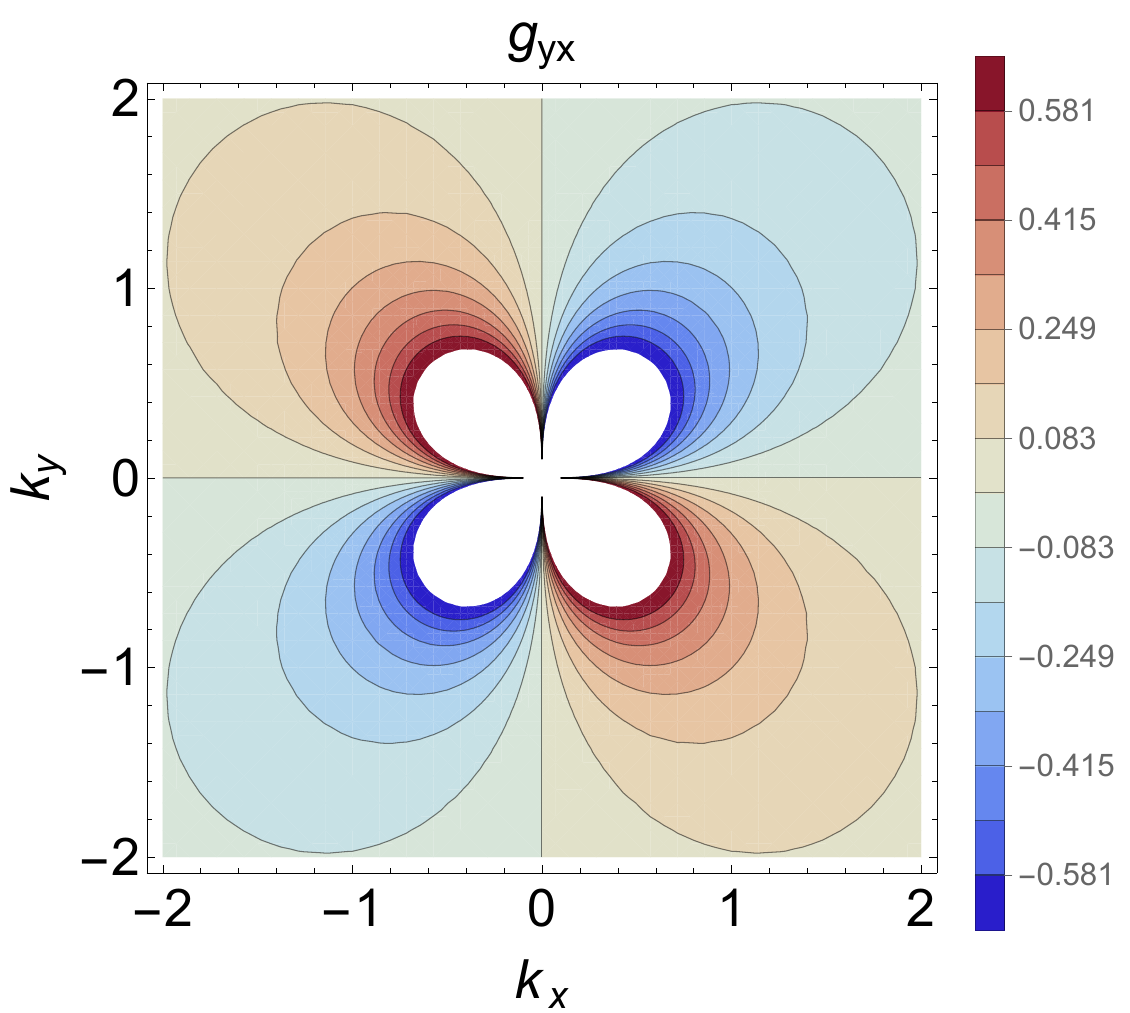}
\caption{
The quantum metric is determined by the derivatives of the texture $\hat{\bm h}$, namely, $g_{\mu \beta} \propto \partial_\mu \hat{\bm h}\cdot \partial_{\beta} \hat{\bm h}$. The components relevant for the present calculation are $g_{xx}$ and $g_{yx}$, whose momentum-space distributions are shown as contour plots. 
While $g_{yx}$ averages to zero due to its parity, it yields a finite contribution to the orbital Hall response when combined with the factor $\mathcal{C}_{\nu,\alpha}$, which compensates its symmetry.}
\label{fig:Lmodel}
\end{figure}

The orbital Hall effect is evaluated within linear response theory.  
Following Ref.~\onlinecite{Bhowal2021}, the orbital current operator is defined as
$ \mathcal{J}^{\gamma}_{\alpha}
= -\frac{\hbar}{2 g_L\mu_B}\{ \mathcal{M}_{\gamma}, v_{\alpha}\}$, where $\mu_B$ is the Bohr magneton, $g_L$ is the Landé $g$ factor, and $v_{\alpha}$ is the velocity operator.
Using the formalism developed in Refs.~\onlinecite{Culcer2017,Burgos2022}, we obtain the orbital Hall response
\begin{align}
\label{Eq:orbitalHall}
\sigma^{\gamma, Hall}_{\alpha,\beta}
= 
\frac{e^2}{8}\epsilon_{\gamma \mu \nu} 
\frac{\hbar}{g_L\mu_B} 
\sum_{\bm k} 
\left( \partial_\mu \hat{\bm h}\cdot \partial_{\beta} \hat{\bm h} \right)  
\mathcal{C}_{\nu,\alpha}(\bm k),
\end{align}
with
\begin{align}
\mathcal{C}_{\nu,\alpha}(\bm k) 
&= \sum_{m,m'(m\neq m')} 
\frac{ \big( v^{mm}_{\bm k,\alpha} + v^{m'm'}_{\bm k,\alpha} \big)
      \big( v^{mm}_{\bm k,\nu}  + v^{m'm'}_{\bm k,\nu} \big) }
     { \varepsilon_{m\bm k} - \varepsilon_{m'\bm k} } \nonumber \\
&\times  [S(S+1)-m^2] n_{FD}(\varepsilon_{m\bm k}),
\end{align}
and $S=3/2$. Eq.~\eqref{Eq:orbitalHall} reveals that the orbital Hall response is entirely governed by the geometry of the degenerate bands.
The factor $\partial_\mu \hat{\bm h} \cdot \partial_\beta \hat{\bm h}$ corresponds to the quantum metric of the underlying SU(2) texture, i.e., the real part of the quantum geometric tensor, which measures the distance between neighboring quantum states in momentum space \cite{YuJiabin2025}. 
In Fig.\eqref{fig:Lmodel} we show the profile of the relevant components on the plane. More precisely, the standard quantum metric is proportional to
$\partial_\mu\hat{\bm h} \cdot  \partial_\beta\hat{\bm h}$.
For the present spin-$3/2$ representation, the proportionality factor is $1/4$, although this prefactor is not universal and generally depends on the total angular momentum and the magnetic quantum number, as shown in Ref.~\cite{BurgosPRB2026}. Accordingly, This numerical factor is absorbed into the overall prefactor of the orbital Hall conductivity.

The resulting Hall response is therefore of purely geometric origin and persists even in the absence of Berry-curvature contributions.
After straightforward algebra, we obtain the explicit analytical expression 
\begin{align}
\sigma^{z,Hall}_{yx}
&= 
\frac{e}{2\pi} \frac{1}{6\pi}
(\gamma_{3/2}+\gamma_{1/2} )^2
\frac{1}{ \sqrt{2}\hbar} \nonumber \\
&\times 
\left( \frac{  3 \sqrt{m_{HH}}  - 7\sqrt{m_{LH}} }{ \gamma_{3/2}-\gamma_{1/2} }  \right)
\sqrt{\varepsilon_F},
\end{align}
where $\gamma_m= \left[\left(\gamma_1+\frac{5}{2}\gamma_2 \right) - 2\gamma_2m^2 \right]$ with $m=3/2$ or $1/2$ for HH or LH respectively. Also, $m_{HH}=m_0/\gamma_{3/2}$ and similarly for $m_{LH}$. 
The magnitude of the response is controlled by the separation between heavy- and light-hole bands, encoded in $(\gamma_{3/2} - \gamma_{1/2})^{-1}$, highlighting its interband origin.
The scaling $\sigma^{z,Hall}_{yx} \propto \sqrt{\varepsilon_F}$ implies linear growth with the Fermi momentum.
Since the present work addresses the intrinsic clean-limit contribution, the quantum-metric-driven orbital Hall response is expected to survive weak disorder. A quantitative analysis of disorder-induced corrections, however, lies beyond the scope of the present work.
In contrast to conventional orbital Hall effects driven by Berry curvature, the present response controlled by the quantum metric probes a complementary sector of quantum geometry, so far overlooked.

\begin{table}[ht]
\centering
\resizebox{\columnwidth}{!}{%
\begin{tabular}{c c c c c c c}
\hline \hline
Material & $\gamma_{1}$ & $\gamma_{2}$ & $\gamma_3$ & $m_{HH}/m_0$ & $m_{LH}/m_0$ & $\sigma^{z,\mathrm{Hall}}_{yx}$ \\
\hline \hline
Si   & 4.29 & 0.34       & 1.45 & 0.401 & 0.164 & $1.013\times10^{4}$ \\
Ge   & 13.38 & 4.24   & 5.69 &0.289& 0.0469 & $-0.313\times10^{4}$ \\
GaAs & 6.85 & 2.10   & 2.90 & 0.526 & 0.084 & $-0.147\times10^{4}$ \\
InSb & 37.17 & 16.50 & 17.79 & 0.471 & 0.0141 & $-3.986\times10^{4}$ \\
\hline \hline
\end{tabular}
}
\caption{Estimated orbital Hall conductivity for representative Luttinger semiconductors with $\varepsilon_F=10$ meV. The effective masses are expressed relative to the free electron mass $m_0$. The orbital Hall conductivity is written in units of $(\hbar/e)$($\Omega$\,m)$^{-1}$.  The spherical approximation is more
accurate when $\gamma_2 \approx \gamma_3$, hence Ge, GaAs and InSb are good candidates to explore the phenomena.}
\label{Eq:tab}
\end{table}

In Table~\eqref{Eq:tab} we provide explicit estimates of the orbital Hall conductivity for four representative semiconductors, using material parameters taken from the literature \cite{Winkler2003}. The resulting magnitudes are comparable (or even larger) to intrinsic orbital Hall conductivities obtained from first-principles calculations in transition metals~\cite{Kontani2008,Kontani2009,JoDaegeun2018,Rang2025}, of the same order as the orbital/spin Hall conductivities reported in semiconductors~\cite{Baek2021,Murakami2003,Cullen2026} and 
comparable to experimental results on metallic heterostructure \cite{GiacomoSala2022} and in Ge \cite{Santos2024}.

The sign of the orbital Hall response is governed by the subtle geometric interplay between heavy-hole and light-hole contributions. In particular, the factor
$\left( \frac{  3 \sqrt{m_{HH}}  - 7\sqrt{m_{LH}} }{ \gamma_{3/2}-\gamma_{1/2} }  \right)$
together with the material-dependent renormalization of the effective masses, naturally accounts for the variation of the sign among different compounds. 
Remarkably, the theory predicts the largest orbital Hall conductivity for InSb, where the isotropic approximation is more accurate.

While the spherical approximation employed here is quantitatively most accurate for materials such as InSb and Ge, and less precise for others, it captures the essential symmetry and geometric structure of the valence bands in a broad class of cubic semiconductors. It therefore provides a minimal and analytically tractable framework for revealing the intrinsic, geometry-driven mechanisms underlying orbital transport.

In realistic semiconductors, the spherical approximation may be relaxed by crystal-field anisotropies or other symmetry-preserving corrections that modify the helicity structure of the valence bands while preserving the combined $\mathcal{I}\Theta$-symmetry. In such cases, the twofold degeneracy remains protected, and the quantum-metric-driven mechanism discussed here continues to provide the intrinsic geometric contribution to the orbital Hall response, although its quantitative value may be modified. By contrast, perturbations that break the protecting symmetries, such as inversion or time-reversal symmetry, lift the degeneracy and may generate additional contributions associated with a finite Berry curvature. The present theory therefore describes the ideal $\mathcal{I}\Theta$-symmetric degenerate limit and identifies the geometric mechanism that survives whenever this symmetry protection is maintained.


Interestingly, the same spherical Luttinger Hamiltonian has recently been employed to study orbital relaxation and multipolar dynamics in $t_{2g}$ systems \cite{Manchon2026}. While that work addresses nonequilibrium relaxation in the presence of disorder, our results demonstrate that, already in the clean limit, the intrinsic orbital Hall response is governed by the off-diagonal matrix structure of the orbital magnetic moment. 
Our results complement these recent developments by showing the fundamentally matrix-valued nature of orbital dynamics in crystalline solids.

\textit{Conclusion.–} 
We have shown that in rotationally invariant spin-$3/2$ systems the orbital Hall response arises entirely from interband processes mediated by the OMM. 
Remarkably, this response is governed by the quantum metric of the bands rather than by the Berry curvature, which vanishes identically in the bulk. 
This intrinsic mechanism is robust against weak disorder and depends solely on the symmetry and band structure of the system, making it broadly applicable to multiband materials with spin-orbit coupling.

Our results suggest a possible route toward disentangling bulk and surface contributions in semiconductor materials. In particular, it is conceivable that Berry-curvature-driven mechanisms dominate nondegenerate surface states, whereas the quantum-metric contribution identified here governs the bulk response in the ideal degenerate limit. 
More generally, our work identifies the off-diagonal orbital magnetic moment as the fundamental microscopic origin of intrinsic orbital Hall transport in degenerate spin-3/2 systems.

\begin{acknowledgments}
We acknowledge support by the Italian Ministry
of Foreign Affairs and International Cooperation, grant
PGR12351 “ULTRAQMAT” and the PNRR MUR project PE0000023-NQSTI.
\end{acknowledgments}

\bibliography{Lmodel}

\end{document}